\def \cmsq           {\hbox{cm$^{-2}$}}
\def \etal         {{\it et~al.} }

\def \kms          {\rm{\hbox{km s$^{-1}$}}}
\def \lam          {$\lambda$}
\def \Lya          {\hbox{Ly$\alpha$}}

\def \zaz          {{$z_a\kern -1.5pt \approx\kern -1.5pt z_e$}}
\def \zllz         {{$z_a\kern -3pt \ll\kern -3pt z_e$}}

\def \zgz          {{\kiA z\lower 3pt \hbox{a} $>$ z\lower 3pt \hbox{e}\ }}
\def \Msun         {\rm{\hbox{M$_{\odot}$}}}           
\documentstyle[11pt,paspconf]{article}

\begin{document}

\title{Elemental Abundances from Intrinsic QSO Emission and Absorption Lines}
\author{F. Hamann}
\affil{Center for Astrophysics \& Space Sciences, University of California 
-- San Diego, La Jolla, CA, 92093-0424}

\begin{abstract}
Several studies have shown that the column densities 
inferred from broad absorption lines (BALs) require extremely 
high metallicities and phosphorus overabundances -- 
apparently in conflict with other abundance diagnostics.  
Here I use {\it HST} spectroscopy of the 
BALQSO PG~1254+047 to argue that the BALs abundance 
estimates are incorrect, because partial line-of-sight coverage of the 
continuum source(s) has led to gross underestimates of the line 
optical depths and column densities. I claim that the significant presence 
of P~V~\lam 1118,1128 absorption in this and other BALQSOs identifies 
the saturated absorption-line spectrum. This interpretation implies 
that the total column densities are at least ten times larger 
than previous estimates, namely log~$N_H$~(\cmsq )~$\ga 22.0$. 
The outflowing BAL gas, at velocities from 
$-$15,000 to $-$27,000~\kms\ in PG~1254+047, is therefore a strong 
candidate for the X-ray absorber in BALQSOs. If this high-column 
density outflow is radiately accelerated, it must 
originate $\la$0.1~pc from the QSO. 

\end{abstract}


\section{Introduction}

Measurements of the elemental abundances in QSOs are important 
for understanding the physics and observable properties 
of the various emission/absorption regions. They 
can also provide valuable constrains on the extent and epoch 
of star formation in galactic nuclei. These constraints could 
be important complements to other studies of high-redshift 
galaxies (involving, for example, the ``Lyman-break'' galaxies or 
damped-Ly$\alpha$ absorbers) that probe more extended structures 
and/or rely on very different data and techniques. 

Three general, independent probes of QSO abundances are 
readily observable at all redshifts: the broad emission lines (BELs), the 
broad absorption lines (BALs) and the narrow associated (intrinsic) 
absorption lines (AALs). Each of these probes has its own 
theoretical and observational uncertainties, some of which might 
be unknown. It is therefore essential to consider as 
many abundance diagnostics as possible. I am now involved in 
several projects to examine a wide range of emission and absorption 
diagnostics in QSOs at different redshifts and luminosities. 
My principle collaborators are Drs. T. Barlow, F. Chaffee, 
G. Ferland, C. Foltz, V. Junkkarinen, K. Korista and J. Shields. 

The most recent work on BEL abundances, based on 
the NV/HeII and NV/CIV line ratios (Hamann \& Ferland 1992, 
1993, Ferland \etal 1996), indicates that the metallicities 
are typically solar or higher and nitrogen is selectively 
enhanced. There is also a significant trend in these line ratios 
suggesting higher metallicities in more luminous QSOs 
(see also Osmer \etal 1994 and Korista \etal 1998). 
Recent AAL studies based on 
high-resolution spectroscopy are in general agreement with the 
BELs, indicating typical metallicities from 
$\sim$1/2 to a few times solar  (see Petitjean \etal 1994, 
Hamann 1997, Tripp \etal 1996, 1997 and refs. therein). 

The most surprising abundance results have come from studies 
of the BALs, where the strengths of the metal-lines 
compared to HI \Lya\ seem to require metallicities (for example 
Si/H) from 20 to $>$100 times solar (Turnshek \etal 1996; 
Hamann 1997). The secure 
detection of broad PV~\lam\lam 1118,1128 absorption 
in one BALQSO, PG~0946+301 (Junkkarinen \etal 1997), 
and tentative detections in two others (Turnshek 1988; 
Korista \etal 1992), suggest further that  phosphorus is highly 
overabundant, with P/C~$>$~60 and P/H~$>$1000 times solar  
(see also Hamann 1997). These abundances, particularly the high 
P/C, are not only in conflict with the other diagnostics but 
they are also incompatible with any enrichment scheme 
dominated by Types I or II supernovae or CNO-processed material 
from stellar envelopes. One possible explanation, suggested 
by Shields (1996), is that the BAL gas is selectively enriched by 
novae. Another possibility is that the BAL abundance estimates 
are incorrect. In this contribution I argue for the latter; 
the PV line has significant strength {\it not} because phosphorus is 
overabundant but because the stronger transitions are more 
optically thick than they appear. The true BAL abundances 
are unknown.

\section{General Absorption Line Analysis}

Absorption lines are, in principle, better abundance 
probes than the emission lines because their strengths 
relate directly to the ionic column densities along our 
line-of-sight. The abundance ratios for any two elements 
$a$ and $b$ simply scale with the column densities,  
\begin{eqnarray}
\left[{a\over b}\right]  = \
\log\left({{N(a_i)}\over{N(b_j)}}\right)  +
\log\left({{f(b_j)}\over{f(a_i)}}\right)  +
\log\left({b\over a}\right)_{\odot} 
\end{eqnarray}
where $(b/a)_{\odot}$ is the solar abundance ratio, and $N$ and $f$ 
are respectively the column densities and ionization fractions 
of elements $a$ and $b$ in ion stages $i$ and $j$. 
The corrections factors, $f(b_j)/f(a_i)$, are often uncertain because 
they depend on the poorly known 
ionization state of the absorber. Hamann (1997) calculated 
ionization fractions and correction factors for a wide range 
of one-zone absorbers in photoionization equilibrium with 
the QSO spectrum. Those calculations showed that the correction 
factors needed to derive metal-to-hydrogen abundance ratios, [M/H], 
from the column densities in HI and a metal ion M$_i$ 
have well-defined minima (at some particular level of ionization). 
Therefore, even if the absorber is 
complex and the ionization state(s) is(are) unknown, we can still 
use the minimum correction factors to place firm lower limits on 
the [M/H] ratios (see Hamann 1997 for details). 
The correction factors for some metal-to-metal 
abundance ratios also have well defined limits that provide 
further robust constraints. One 
example is the correction factor $f$(CIV)/$f$(PV), 
whose minimum value provides lower limits on [P/C]. 

\section{PV and the Problem with BAL Abundance Estimates}

Hamann (1998) recently examined the ionization, column 
densities and metal abundances in the BALQSO 
PG~1254+047. The BAL troughs in this object 
are detached, with velocities ranging from 
$-$15,000 to $-$27,000~\kms , but the spectrum (Fig. 1) 
appears in every other way typical of BALQSOs (cf. Weymann \etal 
1991). Like most optically selected BALQSOs, 
this object has strong BALs in high-ionization species such as  
CIV, NV and OVI without absorption in MgII and other 
low-ionization lines. What makes this source particularly useful 
is its moderate redshift ($z_{em}\approx 1.01$), which allows 
us to measure the BALs at $\lambda \la 1200$~\AA\ in a relatively 
sparse ``\Lya\ forest.'' For example, broad absorption in the 
PV~\lam\lam 1118,1128 doublet is unambiguously detected 
at the same outflow velocities as the stronger 
BALs (Fig. 1; see Hamann 1998 for a full discussion of the 
line identifications). The rarity of previous PV 
detections is probably due to the observational difficulties 
-- namely, complex line blending or low 
signal-to-noise ratios across the PV wavelengths. (To my 
knowledge, there are no existing spectra of BALQSOs that 
can rule out PV absorption at a strength comparable 
to PG~1254+047.) Significant PV absorption might therefore 
be common in BALQSOs. 

\begin{figure}[h]
\plotfiddle{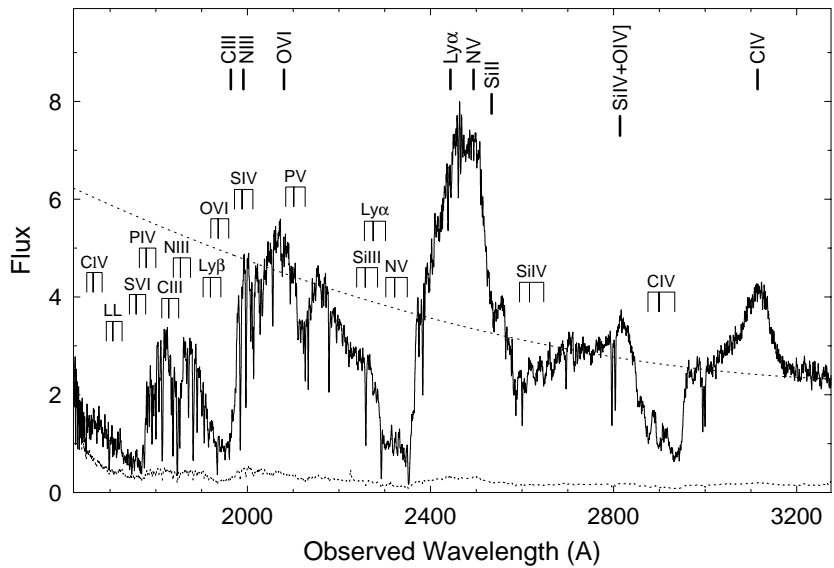}{2.7in}{0.0}{65.0}{65.0}{-200.0}{-190.0}
\caption{{\it HST} spectrum of the BALQSO, PG~1254+047. 
The Flux has units $10^{-15}$ ergs s$^{-1}$ cm$^{-2}$ 
\AA$^{-1}$. The emission lines are labeled 
at $z_{em} = 1.010$ across the top. The BALs are labeled at 3 
redshifts corresponding to the 3 deepest minima in the CIV 
trough. The smooth dotted curve is a simple power-law 
fit to the continuum.}
\end{figure}

Significant PV absorption is, in any case, surprising because 
phosphorus is nominally a rare element; in the sun 
the P/C abundance is $\sim$0.001 while P/O is $\sim$0.0004. 
Hamann (1998) used the analysis outlined in \S2 
to derive specific abundance constraints for PG~1254+047. 
A critical assumption in that analysis is that the measured 
BAL troughs accurately represent the run of optical depth 
versus velocity for each ion. With that assumption, the column 
densities follow by simple integration of the optical depth 
profiles. These column densities, combined with the minimum 
ionizations corrections (from Hamann 1997), 
imply extreme abundances, [C/H]~$\ga$~1.0, [Si/H]~$\ga$~1.8 
and [P/C]~$\ga$~2.2.

\underline{\it However,} 
Hamann (1998) argued that the PV BAL has a significant 
strength {\it not} because phosphorus is overabundant, but 
because the strong transitions like CIV, NV and OVI are 
much more optically thick than they appear. 
Explicit calculations of the line optical depths assuming solar 
relative abundances show that PV is the first weak line to 
appear as the stronger transitions become 
more saturated. The strength of the PV BAL in PG~1254+047 
implies optical depths of, for example, $\ga$6 in \Lya, $\ga$25 
in CIV and $\ga$80 in OVI for solar relative abundances. 
These results indicate that the column densities derived from 
the measured troughs are gross underestimates and, 
consequently, the true abundances are unknown. 

BALs like CIV, OVI and \Lya\ might be optically thick while 
not reaching zero intensity 
if the absorber covers just part of the continuum source(s).
Furthermore, different optically thick lines 
can have different strengths and profiles if their coverage fractions 
differ. Note that coverage fraction differences between BALs 
(that mimic simple 
optical depth or ionization effects in observed spectra) 
can occur naturally if the absorbing regions have a range of 
ionization states or column densities. There is already direct 
evidence for partial coverage, and sometimes 
different coverage fractions in different lines, from the resolved 
multiplet ratios in the narrow components of some BALs 
(Barlow \& Junkkarinen 1994, Wampler \etal 1995) and 
AALs (Barlow \& Sargent 1997; 
Hamann \etal 1997; Hamann 1997 and refs. therein). 
Partial coverage has also been inferred from 
spectropolarimetry of BALQSOs (Cohen \etal 1995, Goodrich \etal 
1995, Hines \& Wills 1995). We cannot measure the coverage 
fractions from multiplet ratios in 
most BALs, but I claim that the PV line signifies partial  
coverage and large line optical depths. Strong support for 
this interpretation comes from the only known AAL system 
with PV absorption, where the resolved doublets clearly 
indicate large optical depths and partial coverage in lines 
such as CIV, NV and SiIV (Barlow \etal 1998). 

\section{Further Consequences of BAL Saturation}

The calculations by Hamann (1998) also show that the PV 
BAL in PG~1254+047 requires a minimum total column density 
(in HI+HII) of $\log N_{\rm H}$(\cmsq )~$\ga 22.0$ 
if the metallicity is roughly solar. 
The actual column density depends on the unknown ionization 
state (higher ionizations require larger total columns for a 
given PV absorption strength). Figure 2 shows the ranges of 
HI and total column densities consistent with the PG~1254+047 
observations for photoionized clouds with different ionization 
parameters ($U\equiv$ the dimensionless ratio of 
hydrogen-ionizing photon to hydrogen particle densities at the 
illuminated face of the clouds). At each $U$, the 
column densities are bounded below by 
the minimum optical depth of $\tau\ga 0.2$ in PV and 
above by the maximum optical depths 
in undetected low-ionization lines like MgII and AlIII 
(see Hamann 1998). 

The total column densities implied by Figure 2 are 
at least an order of magnitude larger than previous estimates 
from the BALs, but they are consistent with 
the large absorbing columns derived from X-ray observations of 
BALQSOs (Mathur \etal 1995; Green \& Mathur 1996). 
In particular, an X-ray absorber with 
$\log N_{\rm H}$(\cmsq )~$\sim 23$ could produce the observed 
BAL spectrum if $+0.4\la \log U \la +0.7$ in a simple one-zone
medium. The outflowing BAL gas, at velocities from $-$15,000 to 
$-$27,000~\kms\ in PG~1254+047, is therefore a strong candidate 
for the X-ray absorber in BALQSOs. 

\begin{figure}[h]
\plotfiddle{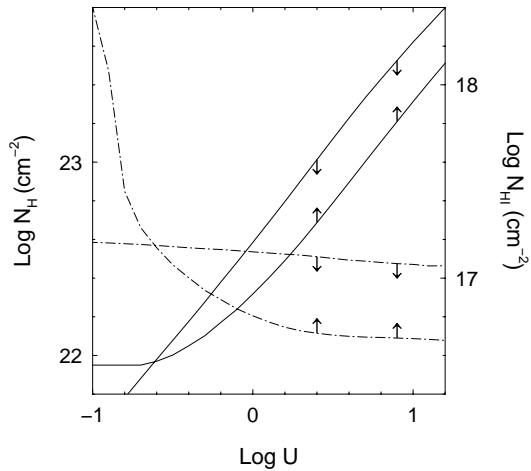}{2.2in}{0.0}{45.0}{45.0}{-150.0}{-124.0}
\caption{Theoretical limits on $N_{\rm H}$ (solid curves, left-hand scale) 
and $N_{\rm HI}$ (dashed curves, right-hand scale) in the BAL region 
of PG~1254+047 for different ionization parameters, $U$, 
in photoionized clouds with solar abundances. 
The permitted values of $N_{\rm H}$ and $N_{\rm HI}$ lie between the two 
pairs of curves for $\log U \ga -0.6$, as indicated by the arrows.}
\end{figure}

While resolving an apparent discrepancy between the UV and 
X-ray data, higher column densities in the BAL region 
might cause serious problems for 
radiative acceleration. The terminal velocity of a radiatively driven 
BAL wind can be expressed as follows,
\begin{equation}
v_{\infty} \ \approx \ 32000\ R_{0.1}^{-{1\over 2}} 
\left({{{f_L L_{46}}\over{N_{22}}} - 
0.008 M_8}\right)^{{1}\over{2}} \ \ {\rm \kms}
\end{equation}
where $R_{0.1}$ is the inner wind radius in units of 0.1~pc, 
$L_{46}$ is the QSO luminosity in units of $10^{46}$~ergs~s$^{-1}$, 
$N_{22}$ is the total column density in $10^{22}$~\cmsq , 
$M_8$ is the central black hole mass relative to 10$^8$~\Msun , and 
$f_L$ is the fraction of the spectral energy distribution absorbed 
or scattered in the wind. $L_{46}\sim 1$ is the 
Eddington luminosity for $M_8 = 1$, and $R_{0.1}\sim 1$ is a 
nominal BEL region radius for $L_{46} = 1$ (cf. Peterson 1993). 
This equation holds strictly for 
open geometries, where the photons escaping one location in the 
BALR are not scattered or absorbed in another location. 
The line optical depth calculations by Hamann (1998) indicate 
that $f_L$ could be as large as a few tenths. 
Therefore, the large total column densities 
implied by the PV analysis (Fig. 2) {\it can} be radiatively 
driven to the observed velocities, but there are 
restrictions on the inner wind radius. For example, if the total 
column in outflowing gas is 
$\log N_{\rm H}$(\cmsq )~$\sim 23$, 
the wind must arise from within $\sim$0.01~pc of the QSO. 

\acknowledgments
This work was supported by NASA grant NAG 5-3234.

\newpage
\bigskip
\noindent{\bf Questions:}
\medskip

{\it M. Crenshaw:} What covering factors do you get for the BALQSOs?

{\it F. Hamann:} I claim that strong BALs like \Lya, CIV 
and OVI are very optically thick, so the observed depths of 
the troughs are direct measures of the coverage fractions. 
For example, from Figure 1 I estimate mean coverage fractions of  
$\sim$70\% in CIV and $\la$50\% in \Lya\ (which is not 
clearly detected). 

{\it B. Wilkes:} If, as you suggest, the BAL gas is the X-ray absorber 
and this absorber does not cover completely the continuum source, 
then we would expect to see soft X-ray emission from the uncovered 
part of the continuum source. Do you know if this is consistent 
with the current X-ray data on BALQSOs? 

{\it F. Hamann:} I'm not sure we can perform that consistency check 
because different ions appear to have different coverage fractions.  
If the high-ionization gas is ``fluffier'' (more extended) 
than the low-ionization 
gas, OVII and OVIII, which might dominate the absorption in 
soft X-rays, could have complete coverage while ions like CIV 
are only partial. Furthermore, the UV and X-ray continuum 
sources could be spatially distinct, so coverage fractions derived 
from the UV lines might not be meaningful for the X-rays. 

{\it N. Arav:} Having UV BALs saturated without being black combined 
with strong X-ray absorption might be the signature of scattering 
of photons into the line of sight as opposed to partial coverage. 

{\it F. Hamann:} Perhaps. Note that 
I use the phrase ``partial coverage'' with 
the understanding that the ``uncovered'' continuum flux might  
come from an extended scattering region. 

{\it M. Gaskell:} Could you comment on the \underline{emission} 
we will see from your optically thicker BAL clouds? 

{\it F. Hamann:} I have not calculated emission line 
fluxes from the BAL gas, but 
the typical high-column density BAL region I propose is 
highly ionized and optically thin at the HI Lyman edge (Fig. 2). 
This gas will not contribute significantly to 
low-ionization BELs such as MgII, but there might be 
some emission in high-ionization lines (depending on 
what the actual column densities and global covering factors are). 
In any case, I would expect emission lines from the BAL gas to 
present very broad and shallow profiles, especially in sources 
like PG~1254+047 where the BAL region appears exclusively at high 
outflow velocities (at least along our line of sight).

\end{document}